\documentclass[conference,10pt]{IEEEtran}
\IEEEoverridecommandlockouts

\usepackage[letterpaper,top=0.75in,bottom=0.75in,left=0.625in,right=0.625in,columnsep=0.25in]{geometry}
\usepackage{newtxtext,newtxmath}
\usepackage{cite}
\usepackage{amsmath}
\usepackage{graphicx}
\usepackage{booktabs}
\usepackage{tabularx}
\usepackage{array}
\usepackage{makecell}
\usepackage{multirow}
\usepackage{dblfloatfix}
\usepackage{microtype}
\usepackage[hidelinks]{hyperref}
\hypersetup{
  pdftitle={Prospective Clinical Indication, Post-Hoc Report Leakage, and Fusion Design in Multi-Image Chest Radiograph Classification: A Patient-Clustered Evaluation},
  pdfauthor={Kamran Shahid and Muhammad Munwar Iqbal},
  pdfsubject={Multi-image chest radiograph classification and report-label leakage},
  pdfkeywords={chest radiography, multimodal learning, clinical indication, report-label circularity, weak supervision, multi-image fusion, patient-cluster bootstrap}
}

\graphicspath{{figures/}}
\newcolumntype{Y}{>{\centering\arraybackslash}X}
\newcolumntype{L}{>{\raggedright\arraybackslash}X}
\newcommand{\tblfont}{\fontsize{8}{9}\selectfont}

\begin{document}

\title{\fontsize{24}{26}\selectfont
Prospective Clinical Indication, Post-Hoc Report Leakage, and Fusion Design in Multi-Image Chest Radiograph Classification: A Patient-Clustered Evaluation}

\author{%
\IEEEauthorblockN{\fontsize{11}{12}\selectfont
Kamran Shahid,
Muhammad Munwar Iqbal\textsuperscript{*}}
\IEEEauthorblockA{\fontsize{10}{11}\selectfont\textit{%
Department of Computer Science, University of Engineering and Technology, Taxila, Pakistan}}
\thanks{Kamran Shahid and Muhammad Munwar Iqbal are with the Department of Computer Science, University of Engineering and Technology, Taxila, Pakistan. Muhammad Munwar Iqbal is an Assistant Professor and the corresponding author. (E-mails: 23-CS-77@students.uettaxila.edu.pk; munwar.iq@uettaxila.edu.pk.)}
}

\maketitle

\begin{abstract}
\bfseries
Chest radiograph datasets often combine multiple images with Clinical Indication, Findings, and Impression, although these inputs are produced at different stages of care. Indication is available before interpretation, whereas Findings and Impression are written after image review. When report-derived labels are predicted from the same post-hoc text, high performance may reflect access to the label source rather than independent image interpretation. We studied 15,000 ReXGradient-160K examinations with two readable images each and evaluated five CheXbert-derived observations under U-Ones, with U-Zeros as a sensitivity endpoint. Frozen DenseNet-121 and Bio+ClinicalBERT encoders produced image and text embeddings. Prospective models used one image, two images, Indication, or multimodal combinations, including ordinary fusion, random-swap training, DeepSets, and SectionGuard-MI. Findings and Impression were used only as leakage controls. Models were trained with five seeds, and public-test uncertainty was estimated with 2,000 patient-cluster bootstrap replicates. Under U-Ones, macro AUROC was 0.643 for the primary image, 0.694 for two images, 0.749 for Indication, and 0.780 for ordinary two-image-plus-Indication fusion. SectionGuard-MI achieved an AUROC of 0.783 and an AUPRC of 0.260. Relative to ordinary fusion, its paired AUROC difference was 0.0031 (95\% CI, $-0.0042$ to 0.0104; adjusted $p=0.374$), while its AUPRC difference was 0.0289 (95\% CI, 0.0095 to 0.0413; adjusted $p=0.004$). DeepSets had the highest prospective AUROC point estimate (0.787), whereas random-swap fusion had the highest prospective AUPRC point estimate (0.265) and better calibration than SectionGuard-MI. Full report text alone reached an AUROC of 0.979 and an AUPRC of 0.836; AUROC remained above 0.973 after exact or expanded masking. Prospective Clinical Indication was strongly associated with the report-derived targets, but the completed comparisons did not isolate its incremental value over the secondary image. Permutation-aware fusion was competitive, and the report-only results showed substantial report-label circularity.
\end{abstract}


\section{Introduction}
Chest radiograph examinations are often represented as a mixture of image files and clinical text, but these elements are created at different points in the reporting process. Clinical Indication is usually recorded before image interpretation and describes why the examination was requested. Findings and Impression are written after the radiologist has reviewed the images. This temporal distinction matters when multimodal models are evaluated. A model that uses only the radiographs and Indication operates with information that could be available before a report is issued. A model that receives Findings or Impression does not. Treating these inputs as equivalent can therefore blur the boundary between prospective prediction and reconstruction of information already recorded in the report.

Large report-linked resources make this problem especially relevant. ReXGradient-160K contains chest radiographs paired with sectioned free-text reports across official train, validation, and test partitions \cite{rexgradient}, while MIMIC-CXR provides another widely used collection of images and reports \cite{mimiccxr}. Automated report labelers such as CheXbert convert radiology text into structured observations at a scale that would be difficult to achieve through manual annotation \cite{chexbert}. Those labels remain derived from the report, however, and should not be presented as independently adjudicated image findings. VisualCheXbert was developed partly in response to evidence that report-derived labels and image-derived labels can disagree \cite{visualchexbert}. If Findings or Impression are supplied to a model whose targets were extracted from those same sections, the model is given direct access to the linguistic source of the endpoint. Performance under that condition is useful as a leakage diagnostic, but it cannot be interpreted as independent radiographic classification.

\subsection{Motivation}
The intended use considered here is prediction before a radiology report exists. At that stage, the available information is limited to the images and the Clinical Indication. The distinction is not merely conceptual: the completed experiments used five CheXbert-derived report observations and explicitly separated prospective inputs from post-hoc report controls. U-Ones was treated as the primary uncertain-label mapping, with U-Zeros analysed as a separate sensitivity endpoint. This design keeps the target definition visible and avoids implying that either mapping represents a single definitive clinical truth. It also makes the interpretation of text performance more precise. Indication may contain symptoms, prior diagnoses, procedures, or follow-up context that are associated with the eventual report observations, whereas Findings and Impression were written after image review and may restate those observations directly \cite{rexgradient,mimiccxr,chexbert,visualchexbert}.

Multi-image examinations create a second source of uncertainty. The selected cohort retained two readable image files for each of 15,000 studies, but the saved primary and secondary positions were file-order slots rather than verified frontal and lateral view labels. A fixed-order fusion model may therefore learn different functions for the two positions even when the clinical meaning of those positions is not stable. The study addressed this issue by comparing fixed-order models with random image swapping during training and with a permutation-invariant DeepSets model \cite{deepsets}. These alternatives test whether competitive performance can be obtained without assigning permanent clinical meaning to file order. The comparison is relevant to the observed dataset structure, but it does not establish that file order caused prediction error because a dedicated paired test of swapped test-time order was not included in the completed results.

The analysis was organised to answer four related questions without extending the claims beyond the available outputs. It compared the primary image, two images, and prospective Indication under the same frozen-feature framework; evaluated ordinary fusion against a parameter-matched baseline, component ablations, random-swap training, DeepSets, and the proposed gated model; measured the residual predictive signal in Findings and Impression after exact and expanded lexical masking; and repeated the main model evaluation under U-Zeros. Public-test uncertainty was estimated with 2,000 patient-cluster bootstrap replicates because the 1,349 test studies represented 1,136 patients, so study-level resampling would not preserve the observed clustering. The completed comparisons support statements about model-level point estimates and specified paired contrasts. They do not isolate the matched incremental value of Indication relative to a secondary image, and the study does not make claims about independently adjudicated diagnosis, external validity, or clinical deployment.

\section{Related Work}
Large chest radiograph resources such as MIMIC-CXR and ReXGradient-160K pair images with free-text reports and make large-scale multimodal experiments possible \cite{rexgradient,mimiccxr}. CheXbert converts report text into structured observations \cite{chexbert}, but those observations remain derived from the radiologist's report. VisualCheXbert showed that report-derived labels and image-based labels can disagree \cite{visualchexbert}. This distinction becomes critical when report text is also supplied as a model input, because strong performance may reflect access to the label source rather than independent interpretation of the radiograph.

Most work on image-text learning in chest radiography has used reports to improve visual representation learning. Zhang et al. learned medical image representations from paired images and text \cite{zhangcontrastive}. GLoRIA aligned global and local visual features with report content \cite{gloria}. Boecking et al. examined how richer report semantics could improve biomedical vision-language processing \cite{boecking}. These studies established the value of paired text during representation learning, but they did not compare prospective Indication with post-hoc report sections as downstream inputs under a common frozen-feature pipeline.

The fusion components examined here also have precedents outside chest radiography. Gated multimodal units learn input-dependent weights for different modalities \cite{gmu}. ModDrop exposes a model to missing modalities during training \cite{moddrop}. Deep Sets provides permutation-invariant aggregation for unordered inputs \cite{deepsets}. These methods are relevant to multi-image radiography because image availability can vary and file order may not encode a stable clinical view. Their use in this study provides controlled alternatives rather than a claim that any individual component is new.

The unresolved issue is experimental. A useful evaluation must separate pre-interpretation text from report sections written after interpretation, compare one- and two-image inputs under the same feature pipeline, test fixed-order fusion against permutation-aware alternatives, and account for repeated patients when estimating uncertainty. The present study was organised around those constraints. Findings and Impression were treated as leakage controls, while prospective models were compared with component ablations and parameter-matched baselines.

\section{Materials and Methods}
\subsection{Data Source and Cohort Construction}
The source dataset was ReXGradient-160K \cite{rexgradient}. The analysis used the official train, validation, and public-test partitions. Candidate studies were obtained from the first two sequential image-archive parts and were required to have two readable image files. Every eligible validation and public-test study exposed by those archive parts was retained. The training candidates were reduced with iterative multilabel stratification \cite{stratification} to obtain exactly 15,000 studies and 30,000 image slots: 12,253 training studies, 1,398 validation studies, and 1,349 public-test studies. The public-test subset represented 1,136 patient clusters. No patient identifier crossed an official split, and no duplicate study key or selected image member was present.

The first-two-part sampling rule was a computational convenience, not a random sample of the complete dataset. The saved representativeness audit compared the selected subset with each full official partition on available variables. Selected studies were younger on average, with standardised mean differences of $-0.239$ in train, $-0.196$ in validation, and $-0.254$ in test. Indications were modestly longer, while Findings and Impression were shorter. Sex distributions were close to the full partitions, with total variation distances of 0.010, 0.036, and 0.008. Normalised source-row position did not show a detectable shift, but that archive-order proxy cannot exclude selection differences in unobserved site, device, or view characteristics. Institution, manufacturer, and ethnicity were unavailable or noninformative in the normalised subset metadata.

\begin{figure*}[!t]
\centering
\includegraphics[width=7.16in]{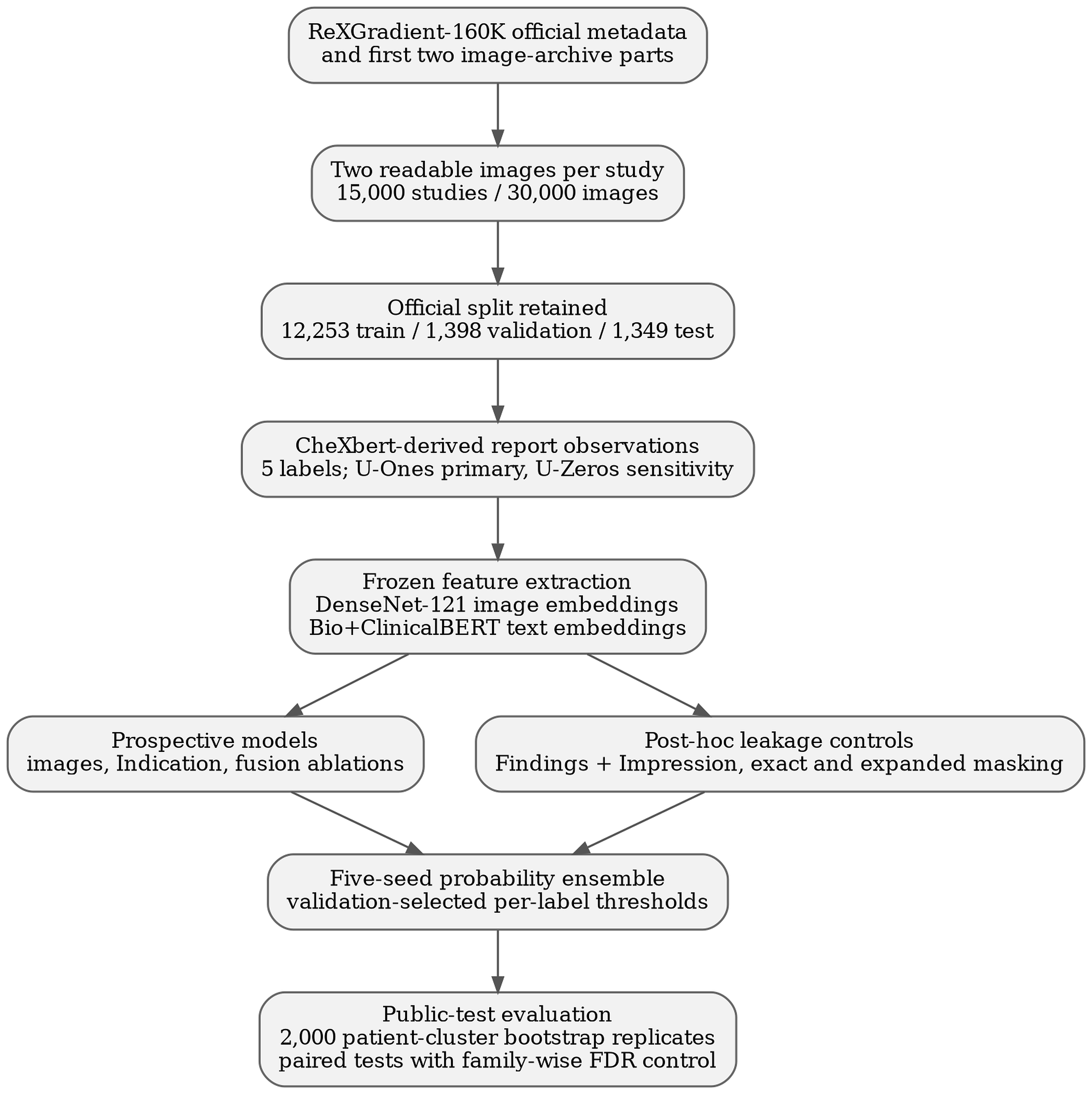}
\caption{Study workflow. The figure distinguishes prospective inputs from post-hoc report controls and shows that model uncertainty was estimated by patient-cluster rather than study-level resampling.}
\label{fig:workflow}
\end{figure*}

\begin{table*}[!t]
\caption{Selected Cohort and U-Ones Positive Study Counts. Percentages Use Studies as the Denominator. The Public-Test Subset Contained 1,136 Unique Patient Clusters.}
\label{tab:cohort}
\centering
\tblfont
\begin{tabularx}{7.16in}{L Y Y Y Y Y Y Y}
\toprule
\textbf{Split} & \textbf{Studies} & \textbf{Images} & \textbf{Atelectasis} & \textbf{Cardiomegaly} & \textbf{Edema} & \textbf{Pleural effusion} & \textbf{Consolidation} \\
\midrule
Train & 12,253 & 24,506 & 1,321 (10.8\%) & 1,387 (11.3\%) & 424 (3.5\%) & 700 (5.7\%) & 268 (2.2\%) \\
Validation & 1,398 & 2,796 & 152 (10.9\%) & 152 (10.9\%) & 59 (4.2\%) & 86 (6.2\%) & 38 (2.7\%) \\
Public test & 1,349 & 2,698 & 144 (10.7\%) & 133 (9.9\%) & 45 (3.3\%) & 66 (4.9\%) & 36 (2.7\%) \\
\bottomrule
\end{tabularx}
\end{table*}

\subsection{Outcome Definition and Uncertainty Policies}
CheXbert \cite{chexbert} was applied to the concatenation of Findings and Impression. Five observations were retained: atelectasis, cardiomegaly, edema, pleural effusion, and consolidation. The primary endpoint used U-Ones, in which positive and uncertain CheXbert outputs were mapped to 1 and negative or unmentioned outputs were mapped to 0. A separate U-Zeros endpoint mapped only explicit positive outputs to 1. These mappings define different report-observation targets and were analysed separately. No result is presented as independently adjudicated radiographic diagnosis.

The public-test U-Ones positive counts were 144 for atelectasis, 133 for cardiomegaly, 45 for edema, 66 for pleural effusion, and 36 for consolidation. Under U-Zeros, the respective counts were 88, 99, 20, 60, and 30. The lower support under U-Zeros is relevant to AUPRC and threshold-dependent metrics, particularly for edema and consolidation.

\subsection{Prospective and Post-Hoc Text Conditions}
Indication was treated as prospective text because it is recorded before interpretation. Findings plus Impression was treated as post-hoc text and used only as a leakage diagnostic. Exact masks removed the five target names. Expanded masks also removed prespecified synonyms and descriptive phrases. The exact report mask changed 8,888 reports and left a mean of 39.03 words; the expanded mask changed 9,698 reports and left a mean of 38.91 words. The complete masking dictionaries are available from the corresponding author upon reasonable request.

The same exact and expanded dictionaries were applied to Indication. Direct target terms were uncommon in the public-test Indications: one affirmed or historical edema mention and six affirmed or historical pleural-effusion mentions were detected. The rule-based assertion categories were used to describe dictionary matches, not as a validated clinical natural-language-processing labeler.

\subsection{Frozen Feature Extraction}
Each image was converted to RGB, resized, and normalised with the standard ImageNet preprocessing pipeline. A frozen ImageNet-pretrained DenseNet-121 \cite{densenet} produced a 1,024-dimensional global-average-pooled feature vector. Indication and report variants were encoded with frozen Bio+ClinicalBERT \cite{clinicalbert}. Indications were truncated at 96 tokens; report variants were truncated at 256 tokens. Mean pooling over non-padding tokens yielded a 768-dimensional text representation. Neither encoder was fine-tuned on ReXGradient-160K.

\subsection{Model Families}
The ordinary fusion baseline multiplied each modality feature by a binary availability indicator, concatenated the available features and indicators, and passed the result to a shared multilabel head. A parameter-matched concatenation model reduced head width so that its trainable parameter count was comparable with SectionGuard-MI. Projection ablations used separate 128-dimensional LayerNorm-linear-ReLU projections for the primary image, secondary image, and text. Additional ablations added gates, whole-modality dropout, a consistency term, or combinations of these components. Random-swap training exchanged the two image positions during training. The DeepSets comparator used a shared image projector and permutation-invariant mean-plus-max aggregation \cite{deepsets}.

For SectionGuard-MI, a missing feature was zeroed before projection, and the projected feature was masked again so that projector bias could not influence the gate network. A joint gate conditioned on all projected features and the availability vector. The gated and masked features, together with the availability indicators, were passed to a shared multilabel head. Unit tests showed a maximum absolute output difference of 0 when absent feature values were perturbed and a maximum difference of 0 when the two images were exchanged in the DeepSets model.

\begin{figure*}[!t]
\centering
\includegraphics[width=7.16in]{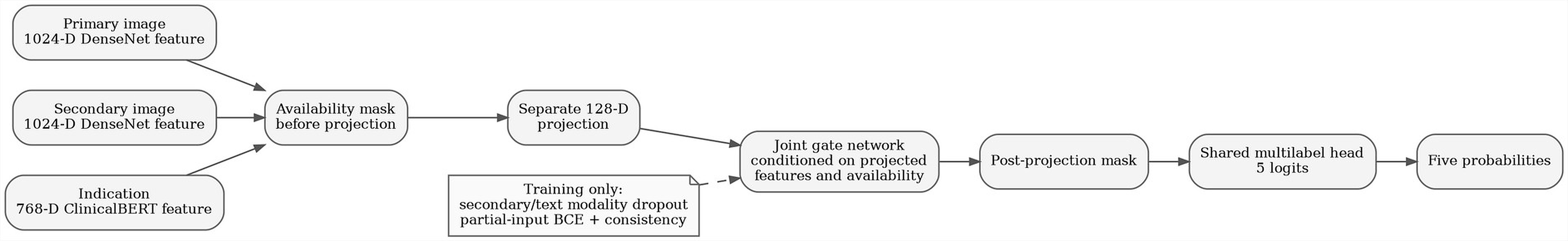}
\caption{SectionGuard-MI feature-level architecture. Missing inputs are masked before and after projection. The gate network receives projected features and availability indicators. Modality dropout and prediction consistency are used only during training.}
\label{fig:architecture}
\end{figure*}

\begin{equation}
\begin{aligned}
\widetilde{h}_m &= a_m h_m, \\
q_m &= a_m\operatorname{ReLU}\!\left(W_m\operatorname{LN}(\widetilde{h}_m)+b_m\right).
\end{aligned}
\label{eq:projection}
\end{equation}

Here, $h_m$ is the frozen embedding for modality $m$ and $a_m$ is its binary availability. For gated models, the concatenated projected features and availability indicators are passed through a two-layer gate network with sigmoid output. The gate output is multiplied by availability before the classification head.

\begin{equation}
\begin{aligned}
g=\sigma\big(&W_{g2}\operatorname{Dropout}[\operatorname{ReLU}(\\
&W_{g1}[q_p;q_s;q_t;a]+b_{g1})]+b_{g2}\big).
\end{aligned}
\label{eq:gate}
\end{equation}

\begin{equation}
\begin{aligned}
z&=[e_pq_p;e_sq_s;e_tq_t;a],\\
\widehat{y}&=\sigma(f_{\mathrm{head}}(z)),\qquad e=g\odot a.
\end{aligned}
\label{eq:head}
\end{equation}

\subsection{Training and Ensemble Inference}
Every model was trained with seeds 42, 1337, 2026, 31415, and 271828. Optimisation used AdamW \cite{adamw} with learning rate $3\times10^{-4}$, weight decay $1\times10^{-4}$, batch size 256, gradient-norm clipping at 5.0, and mixed precision when CUDA was available. Training continued for at most 60 epochs and stopped after 10 epochs without improvement in validation macro AUROC. Positive class weights were the train-set negative-to-positive ratio, floored at 1 and capped at 20.

For SectionGuard-MI, the secondary-image and text branches were independently dropped with probability 0.30 during the partial-input forward pass; the primary image remained present. The loss averaged weighted binary cross-entropy from the full and partial inputs and added a prediction-consistency penalty with weight 0.20. The full-input prediction was stop-gradient in the consistency term.

\begin{equation}
\begin{aligned}
\mathcal{L}={}&\tfrac{1}{2}\mathcal{L}_{\mathrm{BCE}}(y,\ell_{\mathrm{full}})
+\tfrac{1}{2}\mathcal{L}_{\mathrm{BCE}}(y,\ell_{\mathrm{partial}})\\
&+0.20\left\|\sigma(\ell_{\mathrm{partial}})-\operatorname{sg}[\sigma(\ell_{\mathrm{full}})]\right\|_2^2.
\end{aligned}
\label{eq:loss}
\end{equation}

Seed probabilities were combined before thresholding. Validation probabilities were averaged across the five seeds, and one threshold per label was selected to maximise validation F1. Test probabilities were then averaged across seeds and classified with those fixed ensemble thresholds. No threshold was selected on public-test data.

\subsection{Evaluation and Statistical Analysis}
The primary discrimination metrics were macro AUROC and macro AUPRC. Secondary metrics were macro F1, sensitivity, specificity, balanced accuracy, sensitivity at 90\% specificity, macro Brier score, classwise ECE, pooled ECE, micro F1, label accuracy, and exact-match subset accuracy. AUPRC was interpreted with attention to class prevalence because it is more informative than ROC summaries when positive outcomes are rare \cite{prplot}. ECE used ten equal-width probability bins; calibration measures were descriptive and no post-hoc calibration model was included in the completed output set \cite{calibration}.

Confidence intervals used 2,000 patient-cluster bootstrap replicates. Each replicate sampled the 1,136 public-test patients with replacement and included every selected study belonging to each sampled patient. A replicate was rejected if any label contained only one class. Paired comparisons used the same sampled patient clusters for both models. Two-sided bootstrap $p$ values were calculated from the two tail proportions around zero with a plus-one correction. Benjamini-Hochberg adjustment was applied within prespecified exploratory comparison families \cite{bh}. The SectionGuard-MI versus ordinary fusion comparison for macro AUROC and macro AUPRC was the primary family.

\section{Results}
\subsection{Cohort Audit and Endpoint Prevalence}
All five cohort integrity checks passed. The test subset contained 1,349 studies from 1,136 patients, which justified patient-cluster rather than study-level resampling. U-Ones prevalence ranged from 2.7\% for consolidation to 10.7\% for atelectasis in the test subset. The U-Zeros mapping reduced positive support for every label. The representativeness audit also identified a consistent age shift toward younger selected studies and shorter post-hoc report sections relative to the complete official partitions. These differences limit inference from the selected archive subset to the full dataset.

\subsection{Prospective Model Performance}
The primary image alone produced macro AUROC 0.643 (95\% CI, 0.604 to 0.679). Adding a second image raised AUROC to 0.694 (95\% CI, 0.657 to 0.731). Indication alone reached 0.749 (95\% CI, 0.716 to 0.780), which was higher than the two-image point estimate. Ordinary two-image-plus-Indication fusion reached 0.780 (95\% CI, 0.749 to 0.811). These point estimates show a strong predictive association for prospective clinical context, but they do not isolate its incremental contribution relative to the secondary image.

SectionGuard-MI reached macro AUROC 0.783 (95\% CI, 0.751 to 0.813), macro AUPRC 0.260 (95\% CI, 0.217 to 0.321), and macro F1 0.257 (95\% CI, 0.222 to 0.292). DeepSets plus Indication had the largest prospective AUROC point estimate, 0.787 (95\% CI, 0.757 to 0.816). Random-swap projected fusion had the largest prospective AUPRC point estimate, 0.265 (95\% CI, 0.222 to 0.330), and the lowest Brier score and ECE among the prospective fusion models shown in Table~\ref{tab:prospective}. The overlapping confidence intervals do not support a ranking claim based on point estimates alone.

\begin{table*}[!t]
\caption{U-Ones Public-Test Performance for Selected Prospective Models. Confidence Intervals Are Patient-Cluster Bootstrap Intervals Based on 2,000 Replicates. Lower Brier Score and ECE Are Better.}
\label{tab:prospective}
\centering
\tblfont
\begin{tabularx}{7.16in}{L Y Y Y Y Y}
\toprule
\textbf{Model} & \textbf{Macro AUROC (95\% CI)} & \textbf{Macro AUPRC (95\% CI)} & \textbf{Macro F1} & \textbf{Brier} & \textbf{Classwise ECE} \\
\midrule
Primary image & 0.643 (0.604 to 0.679) & 0.117 (0.100 to 0.148) & 0.157 & 0.187 & 0.345 \\
Two images & 0.694 (0.657 to 0.731) & 0.131 (0.112 to 0.164) & 0.191 & 0.183 & 0.329 \\
Indication & 0.749 (0.716 to 0.780) & 0.223 (0.185 to 0.279) & 0.276 & 0.156 & 0.281 \\
Ordinary fusion & 0.780 (0.749 to 0.811) & 0.231 (0.191 to 0.295) & 0.268 & 0.158 & 0.279 \\
Parameter-matched concat. & 0.782 (0.747 to 0.812) & 0.234 (0.196 to 0.298) & 0.280 & 0.158 & 0.283 \\
SectionGuard-MI & 0.783 (0.751 to 0.813) & 0.260 (0.217 to 0.321) & 0.257 & 0.163 & 0.293 \\
Random-swap fusion & 0.785 (0.756 to 0.814) & 0.265 (0.222 to 0.330) & 0.275 & 0.144 & 0.259 \\
DeepSets + Indication & 0.787 (0.757 to 0.816) & 0.261 (0.217 to 0.323) & 0.266 & 0.157 & 0.284 \\
\bottomrule
\end{tabularx}
\end{table*}

\begin{figure*}[!t]
\centering
\includegraphics[width=7.16in]{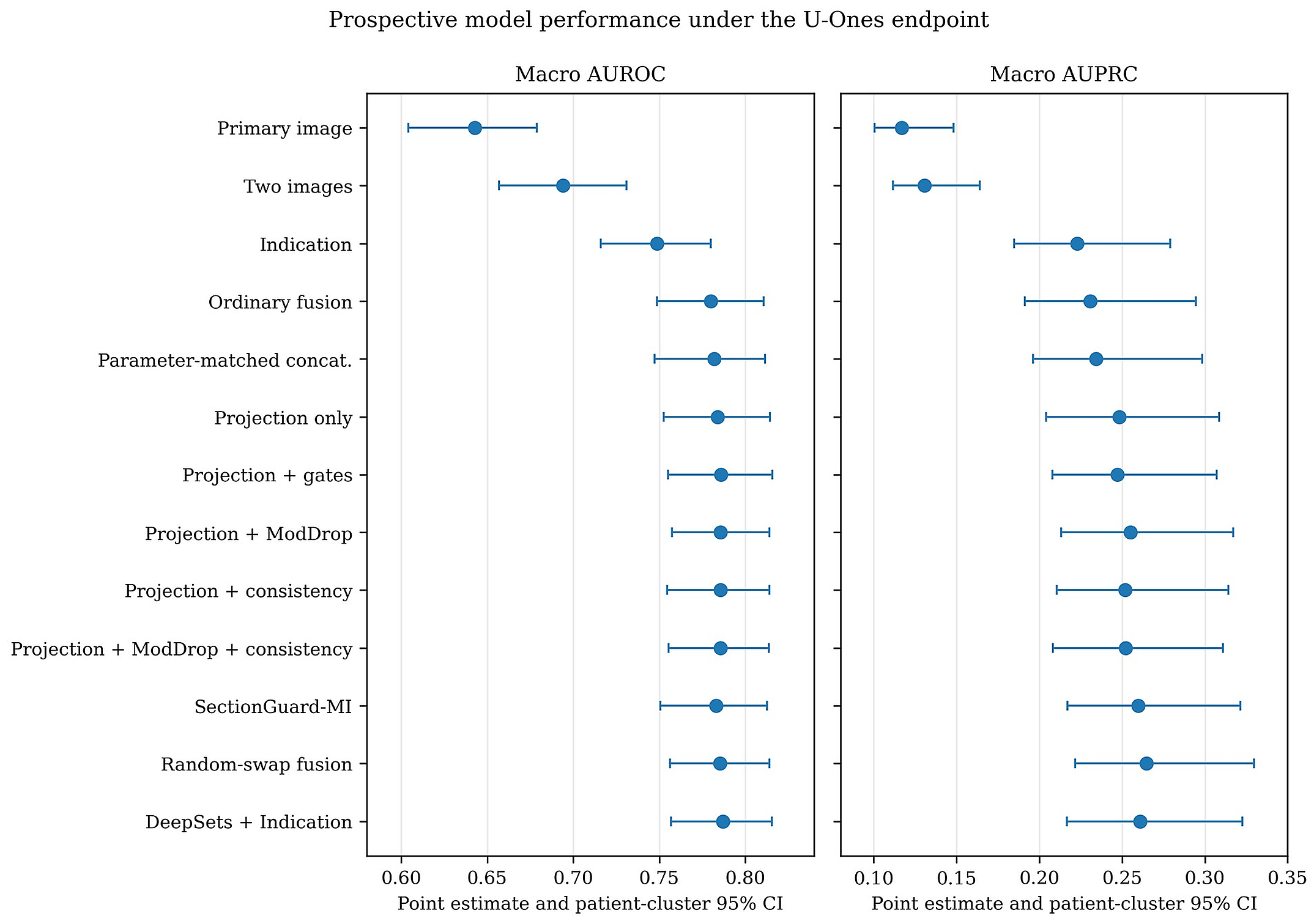}
\caption{Prospective model discrimination under U-Ones. Points are five-seed ensemble estimates; horizontal lines are 95\% patient-cluster bootstrap intervals. Post-hoc report models are excluded from this figure.}
\label{fig:prospective}
\end{figure*}

\subsection{Primary Comparison and Architecture Ablations}
Relative to ordinary two-image-plus-Indication fusion, SectionGuard-MI had a paired macro AUROC difference of 0.0031 (95\% CI, $-0.0042$ to 0.0104; adjusted $p=0.374$). The interval included zero. The paired macro AUPRC difference was 0.0289 (95\% CI, 0.0095 to 0.0413; adjusted $p=0.004$). The results therefore support an AUPRC improvement under the declared endpoint but do not establish an AUROC advantage.

The ablations did not show a consistent additive benefit from all SectionGuard components. Relative to projection only, adding gates did not change AUROC or AUPRC significantly, but it reduced Brier score by 0.0071 and classwise ECE by 0.0036 after false-discovery-rate correction. Adding ModDrop alone did not significantly change discrimination and increased Brier score by 0.0026 and ECE by 0.0093. Adding consistency alone reduced Brier score by 0.0026 without a significant discrimination change. The combination of ModDrop and consistency increased Brier score by 0.0152 and ECE by 0.0297 relative to projection only. These findings indicate that discrimination and calibration responded differently to the same components.

SectionGuard-MI improved AUPRC by 0.0254 relative to parameter-matched concatenation (95\% CI, 0.0048 to 0.0410; adjusted $p=0.027$), but its Brier score was 0.0054 higher and its classwise ECE was 0.0103 higher. DeepSets and random-swap fusion were competitive without the complete gated objective. The trainable parameter counts were 517,902 for SectionGuard-MI, 517,979 for the parameter-matched concatenation model, 334,603 for DeepSets, and 761,099 for ordinary fusion.

\begin{table*}[!t]
\caption{Selected Paired Patient-Cluster Comparisons. Differences Are Model A Minus Model B. Positive Differences Favor Model A for AUROC and AUPRC; Negative Differences Favor Model A for Brier Score and ECE.}
\label{tab:paired}
\centering
\fontsize{7.4}{8.2}\selectfont
\begin{tabularx}{7.16in}{L >{\centering\arraybackslash}p{0.75in} >{\centering\arraybackslash}p{0.68in} >{\centering\arraybackslash}p{1.18in} >{\centering\arraybackslash}p{0.56in} >{\centering\arraybackslash}p{0.63in}}
\toprule
\textbf{Comparison} & \textbf{Metric} & \textbf{Difference} & \textbf{95\% CI} & \textbf{FDR $p$} & \textbf{Reject 0.05} \\
\midrule
SectionGuard-MI minus Ordinary fusion & AUROC & 0.0031 & $-0.0042$ to 0.0104 & 0.3738 & No \\
SectionGuard-MI minus Ordinary fusion & AUPRC & 0.0289 & 0.0095 to 0.0413 & 0.0040 & Yes \\
SectionGuard-MI minus Parameter-matched concat. & AUPRC & 0.0254 & 0.0048 to 0.0410 & 0.0275 & Yes \\
SectionGuard-MI minus Parameter-matched concat. & BRIER & 0.0054 & 0.0034 to 0.0073 & 0.0027 & Yes \\
Projection + gates minus Projection only & BRIER & $-0.0071$ & $-0.0086$ to $-0.0056$ & 0.0027 & Yes \\
Projection + gates minus Projection only & CLASSWISE ECE & $-0.0036$ & $-0.0055$ to $-0.0018$ & 0.0027 & Yes \\
Projection + ModDrop minus Projection only & BRIER & 0.0026 & 0.0014 to 0.0038 & 0.0027 & Yes \\
Projection + consistency minus Projection only & BRIER & $-0.0026$ & $-0.0037$ to $-0.0015$ & 0.0027 & Yes \\
Projection + ModDrop + consistency minus Projection only & BRIER & 0.0152 & 0.0139 to 0.0166 & 0.0027 & Yes \\
Projection + ModDrop + consistency minus Projection only & CLASSWISE ECE & 0.0297 & 0.0281 to 0.0314 & 0.0027 & Yes \\
\bottomrule
\end{tabularx}
\end{table*}

\begin{figure*}[!t]
\centering
\includegraphics[width=7.16in]{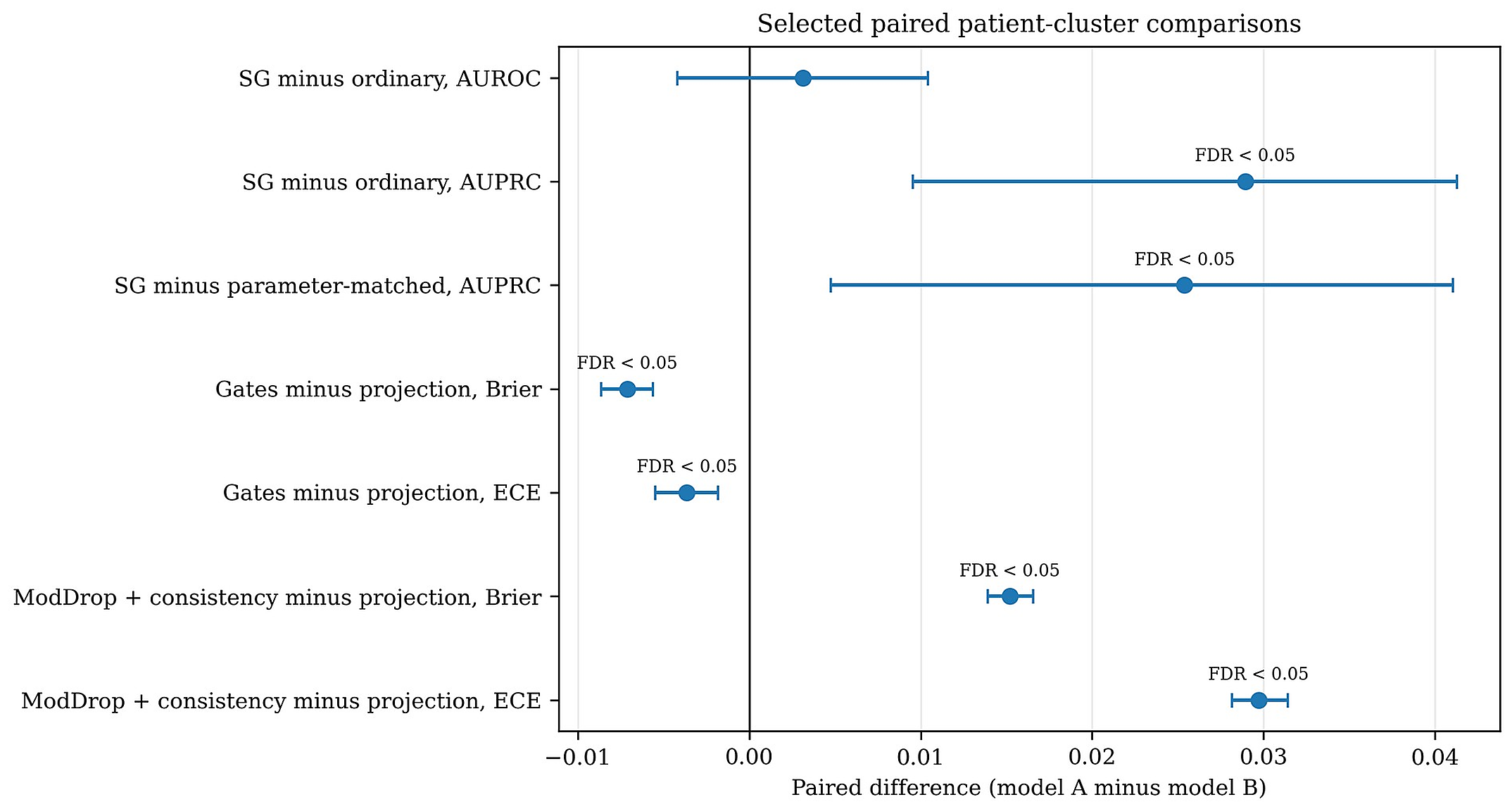}
\caption{Selected paired effects. Confidence intervals are based on the same patient-cluster samples for both models. Calibration metrics have the opposite direction from discrimination metrics: lower values are better.}
\label{fig:paired}
\end{figure*}

\subsection{Image-Order Handling}
The saved cohort used positional primary and secondary image slots rather than validated clinical view labels. Randomly exchanging those positions during training yielded macro AUROC 0.785 and AUPRC 0.265. The permutation-invariant DeepSets model yielded AUROC 0.787 and AUPRC 0.261. Both were at least comparable by point estimate with the fixed-order prospective fusion models. This pattern does not prove that image order caused an error, because a dedicated paired test of swapped test-time order was not present in the completed result package. It does show that a model need not depend on a fixed positional interpretation to achieve the observed performance range.

\subsection{Indication Masking}
Exact masking of target names produced Indication-only AUROC 0.749, compared with 0.749 for the unmasked text. The paired AUROC difference was $-0.0002$ (95\% CI, $-0.0036$ to 0.0033; adjusted $p=0.923$). Expanded masking produced AUROC 0.750; the paired difference from unmasked Indication was $-0.0011$ (95\% CI, $-0.0049$ to 0.0029; adjusted $p=0.758$). AUPRC was 0.223 for unmasked Indication, 0.215 after exact masking, and 0.212 after expanded masking. The unmasked-versus-expanded AUPRC difference was 0.0111 (95\% CI, 0.0022 to 0.0189; adjusted $p=0.050$).

The small AUROC changes are consistent with the low frequency of direct target terms in public-test Indications. They do not establish that Indication is free of outcome-related information. Symptoms, procedures, prior diagnoses, and clinical context can remain predictive without matching the target dictionaries.

\subsection{Post-Hoc Report Leakage and Image-Text Interference}
Full Findings plus Impression alone reached macro AUROC 0.979 (95\% CI, 0.968 to 0.988), macro AUPRC 0.836 (95\% CI, 0.792 to 0.874), and macro F1 0.753. Exact masking reduced AUROC by 0.0054 and AUPRC by 0.0521. Expanded masking reduced AUROC by 0.0059 and AUPRC by 0.0863. All four discrimination differences remained significant after within-family correction. Despite these reductions, the masked report models retained AUROC above 0.973, far above every prospective condition.

The residual performance after masking is compatible with distributed report-label information. Findings and Impression can express location, severity, temporal change, treatment, and descriptive morphology without using the exact target name. The experiment does not identify which phrases carry that information, and the term ``non-lexical leakage'' would be too broad. The supported conclusion is narrower: the tested exact and expanded dictionaries removed only a limited part of the report-derived signal.

Adding two image embeddings to the full report reduced AUROC by 0.0258 and AUPRC by 0.1364 relative to report alone; both differences had adjusted $p=0.001$. A gated image-plus-report model reduced the loss but remained below report alone in AUROC by 0.0068 and in AUPRC by 0.0572. Its Brier score and ECE were lower than those of report alone. Thus, image addition altered discrimination and calibration in different directions. The result is empirical evidence of fusion interference in this feature-level setting, not evidence that images contain negative clinical information.

\begin{table*}[!t]
\caption{Post-Hoc Report Controls. Findings and Impression Were Also Used to Derive the CheXbert Endpoints; These Models Are Leakage Diagnostics and Not Deployable Prospective Classifiers.}
\label{tab:posthoc}
\centering
\tblfont
\begin{tabularx}{7.16in}{L Y Y Y Y Y}
\toprule
\textbf{Post-hoc condition} & \textbf{Macro AUROC (95\% CI)} & \textbf{Macro AUPRC (95\% CI)} & \textbf{Macro F1} & \textbf{Brier} & \textbf{ECE} \\
\midrule
Full report only & 0.979 (0.968 to 0.988) & 0.836 (0.792 to 0.874) & 0.753 & 0.035 & 0.054 \\
Exact-masked report only & 0.974 (0.965 to 0.981) & 0.784 (0.741 to 0.829) & 0.716 & 0.048 & 0.069 \\
Expanded-masked report only & 0.973 (0.965 to 0.981) & 0.749 (0.709 to 0.798) & 0.657 & 0.049 & 0.072 \\
Two images + full report & 0.953 (0.940 to 0.965) & 0.699 (0.656 to 0.747) & 0.627 & 0.037 & 0.042 \\
Two images + exact-masked report & 0.946 (0.933 to 0.958) & 0.642 (0.596 to 0.695) & 0.581 & 0.055 & 0.069 \\
Two images + expanded-masked report & 0.944 (0.932 to 0.956) & 0.614 (0.573 to 0.668) & 0.576 & 0.058 & 0.078 \\
Gated two images + full report & 0.972 (0.961 to 0.982) & 0.779 (0.738 to 0.828) & 0.728 & 0.028 & 0.037 \\
\bottomrule
\end{tabularx}
\end{table*}

\begin{figure*}[!t]
\centering
\includegraphics[width=7.16in]{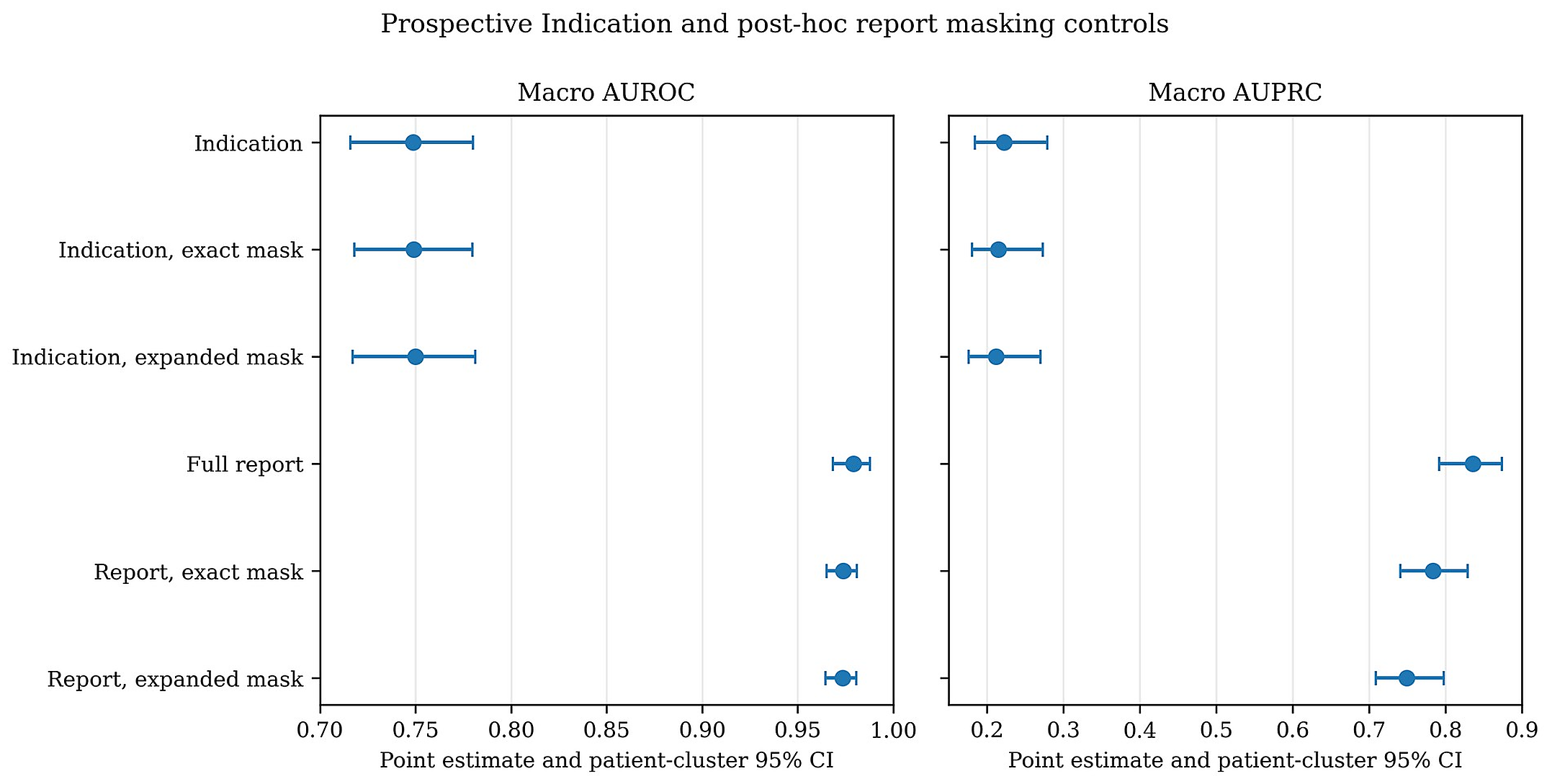}
\caption{Text masking audit. Indication is prospective. Full report denotes Findings plus Impression and is post-hoc. Exact and expanded masks reduce report performance but do not remove the large gap between report-based and prospective conditions.}
\label{fig:masking}
\end{figure*}

\subsection{Per-Label Findings}
Performance varied across labels. For SectionGuard-MI, per-label AUROC ranged from 0.716 for atelectasis to 0.855 for pleural effusion. Its AUPRC ranged from 0.067 for consolidation to 0.415 for pleural effusion. Random-swap and DeepSets models showed a similar pattern. Consolidation had only 36 U-Ones positives in the test subset, and its AUPRC remained low across prospective models. The full report model produced per-label AUROC between 0.956 and 0.997, which is consistent with the report-derived endpoint rather than independent image truth.

\begin{figure*}[!t]
\centering
\includegraphics[width=7.16in]{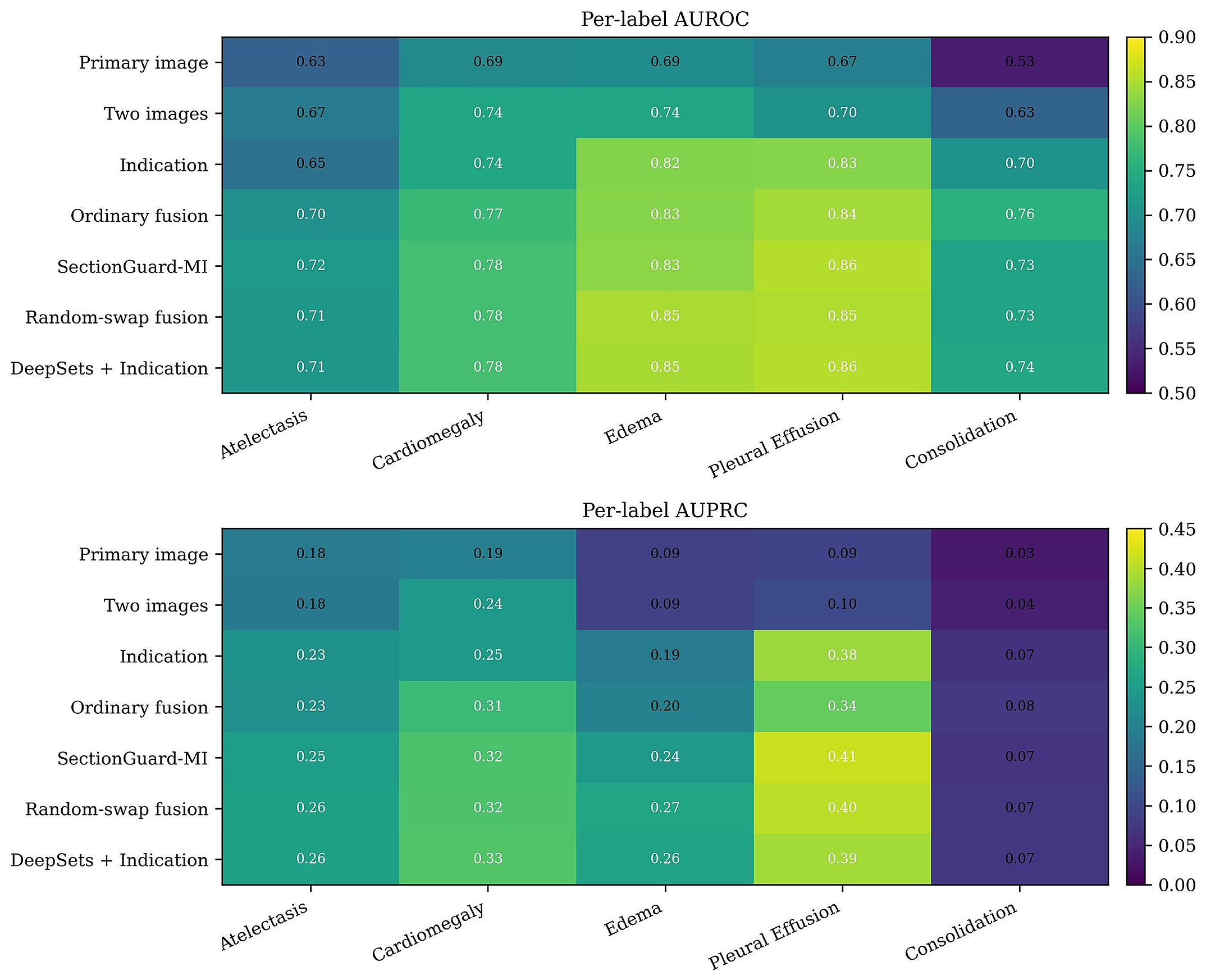}
\caption{Per-label prospective performance under U-Ones. Each cell is a point estimate from the five-seed ensemble. The heatmaps show that label-specific difficulty was not uniform and that low-prevalence consolidation had low AUPRC in every prospective model.}
\label{fig:perlabel}
\end{figure*}

\subsection{U-Zeros Sensitivity}
The U-Zeros endpoint produced the same broad ordering by point estimate. Primary-image and two-image AUROC were 0.651 and 0.700. Indication-only AUROC was 0.755. Ordinary fusion reached 0.785, SectionGuard-MI reached 0.789, and DeepSets reached 0.791. SectionGuard-MI had U-Zeros AUPRC 0.222, ordinary fusion 0.191, and DeepSets 0.216. These results show that the principal distinction between image-only and Indication-informed models was not unique to U-Ones. No paired U-Zeros comparisons were included in the completed result table, so these values are presented as sensitivity point estimates rather than formal significance claims.

\begin{table*}[!t]
\caption{U-Zeros Sensitivity Results. These Are Five-Seed Ensemble Point Estimates. Patient-Cluster Confidence Intervals Are Not Reproduced Here, and Paired U-Zeros Comparisons Were Not Prespecified in the Completed Output Set.}
\label{tab:uzeros}
\centering
\tblfont
\begin{tabularx}{7.16in}{L Y Y Y Y Y}
\toprule
\textbf{Model} & \textbf{Macro AUROC} & \textbf{Macro AUPRC} & \textbf{Macro F1} & \textbf{Brier} & \textbf{Classwise ECE} \\
\midrule
Primary image & 0.651 & 0.083 & 0.117 & 0.162 & 0.319 \\
Two images & 0.700 & 0.096 & 0.146 & 0.158 & 0.298 \\
Indication & 0.755 & 0.190 & 0.249 & 0.136 & 0.259 \\
Ordinary fusion & 0.785 & 0.191 & 0.244 & 0.158 & 0.277 \\
SectionGuard-MI & 0.789 & 0.222 & 0.255 & 0.130 & 0.238 \\
DeepSets + Indication & 0.791 & 0.216 & 0.245 & 0.141 & 0.265 \\
\bottomrule
\end{tabularx}
\end{table*}

\begin{figure*}[!t]
\centering
\includegraphics[width=7.16in]{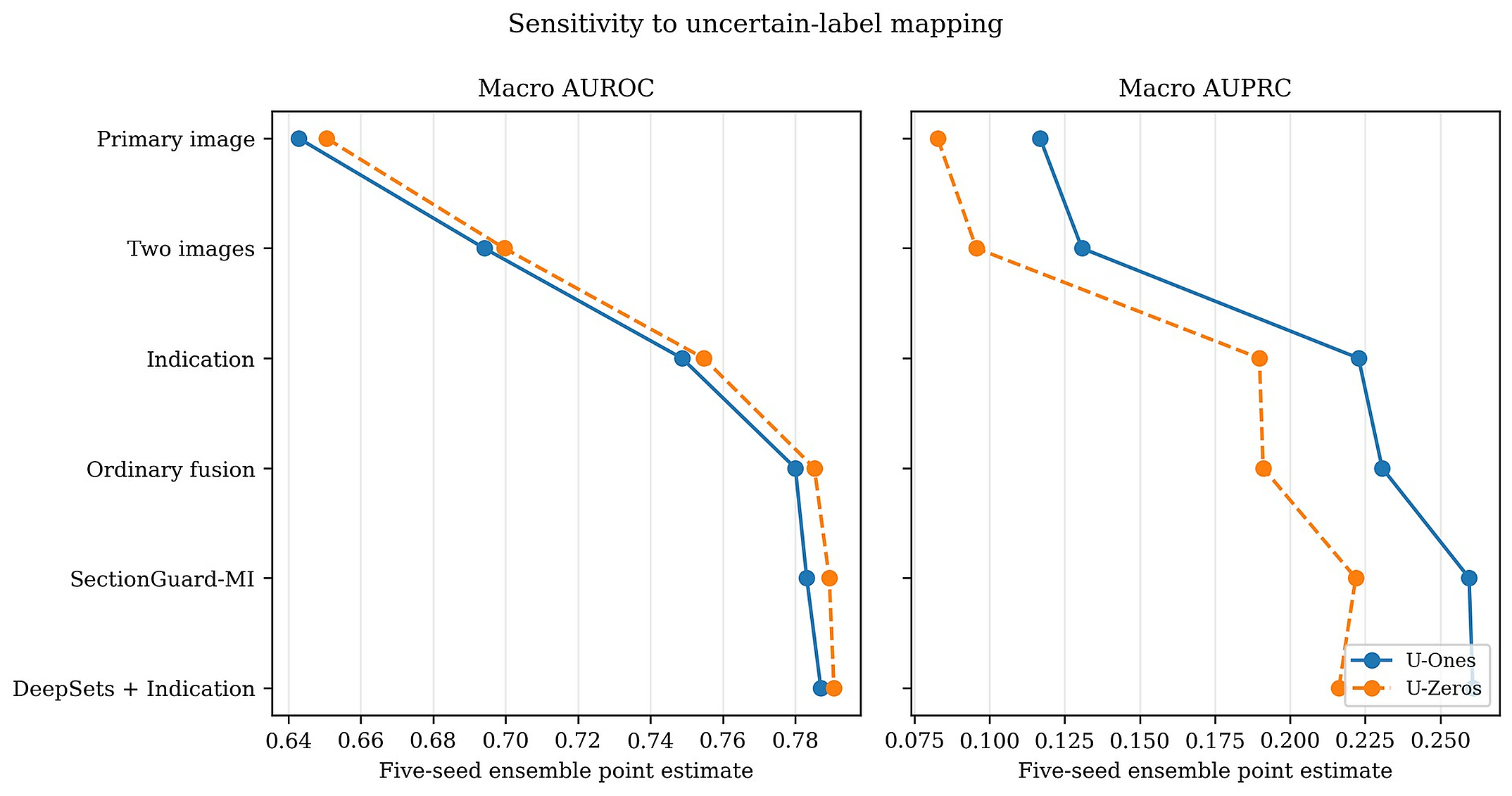}
\caption{U-Ones and U-Zeros sensitivity. The two policies define different report-observation endpoints. Lines connect point estimates only and do not imply paired statistical equivalence.}
\label{fig:sensitivity}
\end{figure*}

\section{Discussion}
The prospective results showed a strong association between Clinical Indication and the five report-derived observations. The Indication-only model had a higher AUROC point estimate than the two-image model, while two-image input improved on the primary image alone. Because the completed analysis did not include matched incremental comparisons of primary image plus Indication against primary plus secondary image, these results do not establish that Indication adds more information than the second image. They show that pre-interpretation clinical context was strongly predictive in this selected cohort.

SectionGuard-MI produced a significant AUPRC improvement over ordinary fusion but not a significant AUROC improvement. Both statements are needed. AUPRC is important for the rare outcomes in this study, and the 0.0289 paired gain was not trivial relative to the ordinary-fusion AUPRC of 0.231. At the same time, the AUROC interval spanned a small benefit and a small disadvantage. The model should not be described as generally superior. Its Brier score and ECE were also worse than those of ordinary fusion, random-swap fusion, and several simpler ablations. A model can rank positive studies reasonably while assigning probabilities that are poorly aligned with observed frequencies \cite{calibration}.

The component analysis explains why a single headline comparison would be inadequate. Gates improved Brier score and ECE relative to projection only without improving discrimination. ModDrop alone and the ModDrop-plus-consistency combination worsened calibration. SectionGuard-MI recovered part of that calibration loss relative to the combined ablation, yet remained less calibrated than the parameter-matched concatenation baseline. These outcomes are plausible because modality dropout changes the training distribution, while consistency regularisation can make partial-input predictions resemble the full-input teacher without guaranteeing calibrated probabilities. The data do not support a stronger mechanistic claim.

Image-order handling also mattered to interpretation. The saved primary and secondary slots were positional, not confirmed frontal and lateral views. Random-swap training and a permutation-invariant DeepSets aggregator achieved competitive results, and DeepSets had the highest prospective AUROC point estimate. This does not establish that DeepSets is the best model; the relevant intervals overlapped and the completed paired table did not compare DeepSets directly with SectionGuard-MI. It does show that future multi-image studies should either validate image semantics or use an architecture that does not assign clinical meaning to file order.

The report controls provide the clearest evidence in the study. Findings and Impression were the source text used to derive the CheXbert observations, and report-only AUROC approached 0.98. Exact and expanded masking produced statistically significant losses, especially in AUPRC, but AUROC remained above 0.97. Keyword removal therefore did not make post-hoc report text comparable with prospective text. Descriptive report content still encodes the observation through anatomy, severity, temporal comparison, and management context. Because independent image annotations were not available, these post-hoc models should be treated only as leakage diagnostics.

The Indication experiment requires a different interpretation. Exact and expanded masking changed AUROC by about one-thousandth, and direct target terms occurred in only seven public-test Indications. The result argues against a simple explanation in which Indication performance is driven mainly by the five target names. It does not exclude subtler forms of target-related context. A patient sent for follow-up of heart failure, for example, may have a high prior probability of edema or pleural effusion even if neither target term appears.

The uncertainty-policy analysis showed that the qualitative separation between image-only and Indication-informed models persisted under U-Zeros. However, fewer positive studies changed AUPRC, F1, and class weights. U-Ones and U-Zeros should not be treated as alternative estimates of a single true target. They are different operational definitions of report uncertainty, and both remain weak labels.

\subsection{Limitations}
The outcomes were extracted from reports and were not independently assigned by radiologists who reviewed the images. Report observations and image findings can disagree \cite{visualchexbert}. The reported numbers therefore quantify prediction of a report-derived endpoint. They cannot establish diagnostic accuracy, clinical utility, or radiologist-level performance.

Cohort selection depended on the first two sequential image-archive parts. The selected studies were younger and had different report-section lengths than the full official partitions. The source-row audit did not identify an archive-position shift, but it cannot test unobserved institution, device, acquisition protocol, or view type. External validity to the complete ReXGradient-160K dataset or another clinical dataset remains uncertain.

The image encoder and text encoder were frozen. This design isolated feature-level fusion and reduced computational demand, but it does not show how end-to-end fine-tuning would alter the ranking of models. Conversely, end-to-end training could introduce additional dataset-specific dependencies that are absent from the present experiment.

The completed result package did not include independent external validation, radiologist adjudication, prospective evaluation, post-hoc probability calibration, or a completed test-time progressive missingness experiment. The manuscript therefore makes no claim about deployment under missing inputs, calibrated clinical risk, or clinical decision support. The public-test set was reused for many exploratory comparisons; false-discovery-rate correction limits but does not eliminate the possibility of chance exploratory findings.

Per-label confidence intervals were not included in the completed tables, and some labels had few positive test studies. The macro estimates give equal weight to each label, but they do not remove the uncertainty caused by 36 consolidation and 45 edema positives under U-Ones. Finally, the direct-term detector and masking dictionaries were transparent rule sets, not validated clinical NLP tools. Residual report performance should be interpreted as failure of those specific masks, not proof that no lexical masking strategy could ever reduce leakage further.

\section{Conclusion}
In a 15,000-study, two-image subset of ReXGradient-160K, the Indication-only model had a higher AUROC point estimate than the two-image model, but the completed comparisons did not isolate its incremental contribution relative to the secondary image. SectionGuard-MI increased AUPRC relative to ordinary fusion but did not establish an AUROC advantage and did not provide the best calibration. Random-swap and permutation-invariant image fusion were competitive, which is relevant when image positions lack validated clinical meaning. Findings and Impression produced near-ceiling performance because the endpoint was extracted from the same post-hoc text; exact and expanded lexical masking reduced that performance but left substantial report-label circularity. Future multimodal chest radiograph studies should separate prospective and post-interpretation text, define the target as report-derived when independent image truth is absent, account for repeated patients in uncertainty estimation, and report calibration and component-matched comparisons alongside discrimination.

\section*{Declarations}
\subsection*{Ethics and Data Governance}
This secondary computational analysis used de-identified ReXGradient-160K data under the provider's access and data-use conditions. No direct patient contact occurred. Because this study relied entirely on a publicly available, completely de-identified dataset, formal institutional review board (IRB) review or exemption determination was not required by the authors' institution.

\subsection*{Data Availability}
ReXGradient-160K is available from its provider under the applicable access conditions \cite{rexgradient}. Raw images and report text are not redistributed with this manuscript. Aggregate result tables may be shared only to the extent permitted by the dataset agreement.

\subsection*{Code Availability}
The analysis code, executed notebook, and derived aggregate result files are available from the corresponding author upon reasonable request. Raw images and report text are not redistributed and remain subject to the ReXGradient-160K data-use conditions.

\subsection*{Funding}
This research received no specific grant from any funding agency in the public, commercial, or not-for-profit sectors.

\subsection*{Competing Interests}
The authors declare that they have no competing interests.

\subsection*{Author Contributions}
Kamran Shahid: Conceptualization, methodology, software, data curation, formal analysis, investigation, visualization, and writing---original draft. Muhammad Munwar Iqbal: Supervision, methodology, validation, and writing---review and editing. Both authors reviewed and approved the final manuscript.

\bibliographystyle{IEEEtran}
\bibliography{references}

\begin{thebibliography}{10}
\providecommand{\url}[1]{#1}
\csname url@samestyle\endcsname
\providecommand{\newblock}{\relax}
\providecommand{\bibinfo}[2]{#2}
\providecommand{\BIBentrySTDinterwordspacing}{\spaceskip=0pt\relax}
\providecommand{\BIBentryALTinterwordstretchfactor}{4}
\providecommand{\BIBentryALTinterwordspacing}{\spaceskip=\fontdimen2\font plus
\BIBentryALTinterwordstretchfactor\fontdimen3\font minus
  \fontdimen4\font\relax}
\providecommand{\BIBforeignlanguage}[2]{{%
\expandafter\ifx\csname l@#1\endcsname\relax
\typeout{** WARNING: IEEEtran.bst: No hyphenation pattern has been}%
\typeout{** loaded for the language `#1'. Using the pattern for}%
\typeout{** the default language instead.}%
\else
\language=\csname l@#1\endcsname
\fi
#2}}
\providecommand{\BIBdecl}{\relax}
\BIBdecl

\bibitem{rexgradient}
X.~Zhang, J.~N. Acosta, J.~Miller, O.~Huang, and P.~Rajpurkar,
  ``{ReXGradient-160K}: A large-scale publicly available dataset of chest
  radiographs with free-text reports,'' arXiv preprint, 2025, doi:
  10.48550/arXiv.2505.00228.

\bibitem{mimiccxr}
A.~E.~W. Johnson, T.~J. Pollard, S.~J. Berkowitz \emph{et~al.}, ``{MIMIC-CXR},
  a de-identified publicly available database of chest radiographs with
  free-text reports,'' \emph{Scientific Data}, vol.~6, p. 317, 2019, doi:
  10.1038/s41597-019-0322-0.

\bibitem{chexbert}
A.~Smit, S.~Jain, P.~Rajpurkar, A.~Pareek, A.~Y. Ng, and M.~P. Lungren,
  ``{CheXbert}: Combining automatic labelers and expert annotations for
  accurate radiology report labeling using {BERT},'' in \emph{Proceedings of
  EMNLP}, 2020, pp. 1500--1519, doi: 10.18653/v1/2020.emnlp-main.117.

\bibitem{visualchexbert}
S.~Jain, A.~Smit, S.~Q.~H. Truong \emph{et~al.}, ``{VisualCheXbert}: Addressing
  the discrepancy between radiology report labels and image labels,'' in
  \emph{Proceedings of the Conference on Health, Inference, and Learning},
  2021, doi: 10.1145/3450439.3451862.

\bibitem{deepsets}
M.~Zaheer, S.~Kottur, S.~Ravanbakhsh, B.~P{\'o}czos, R.~Salakhutdinov, and
  A.~J. Smola, ``Deep sets,'' in \emph{Advances in Neural Information
  Processing Systems}, vol.~30, 2017, doi: 10.5555/3294996.3295098.

\bibitem{zhangcontrastive}
Y.~Zhang, H.~Jiang, Y.~Miura, C.~D. Manning, and C.~P. Langlotz, ``Contrastive
  learning of medical visual representations from paired images and text,'' in
  \emph{Proceedings of Machine Learning for Healthcare}, 2022, doi:
  10.48550/arXiv.2010.00747.

\bibitem{gloria}
S.~C. Huang, L.~Shen, M.~P. Lungren, and S.~Yeung, ``{GLoRIA}: A multimodal
  global-local representation learning framework for label-efficient medical
  image recognition,'' in \emph{Proceedings of ICCV}, 2021, pp. 3942--3951,
  doi: 10.1109/ICCV48922.2021.00391.

\bibitem{boecking}
B.~Boecking, N.~Usuyama, S.~Bannur \emph{et~al.}, ``Making the most of text
  semantics to improve biomedical vision-language processing,'' in
  \emph{European Conference on Computer Vision}, 2022, pp. 1--21, doi:
  10.1007/978-3-031-20059-5\_1.

\bibitem{gmu}
J.~Arevalo, T.~Solorio, M.~M. y~G{\'o}mez, and F.~A. Gonz{\'a}lez, ``Gated
  multimodal units for information fusion,'' arXiv preprint, 2017, doi:
  10.48550/arXiv.1702.01992.

\bibitem{moddrop}
N.~Neverova, C.~Wolf, G.~W. Taylor, and F.~Nebout, ``{ModDrop}: Adaptive
  multi-modal gesture recognition,'' \emph{IEEE Transactions on Pattern
  Analysis and Machine Intelligence}, vol.~38, no.~8, pp. 1692--1706, 2016,
  doi: 10.1109/TPAMI.2015.2461544.

\bibitem{stratification}
K.~Sechidis, G.~Tsoumakas, and I.~Vlahavas, ``On the stratification of
  multi-label data,'' in \emph{Machine Learning and Knowledge Discovery in
  Databases}, 2011, pp. 145--158, doi: 10.1007/978-3-642-23808-6\_10.

\bibitem{densenet}
G.~Huang, Z.~Liu, L.~van~der Maaten, and K.~Q. Weinberger, ``Densely connected
  convolutional networks,'' in \emph{Proceedings of CVPR}, 2017, pp.
  2261--2269, doi: 10.1109/CVPR.2017.243.

\bibitem{clinicalbert}
E.~Alsentzer, J.~R. Murphy, W.~Boag \emph{et~al.}, ``Publicly available
  clinical {BERT} embeddings,'' in \emph{Proceedings of the 2nd Clinical
  Natural Language Processing Workshop}, 2019, pp. 72--78, doi:
  10.18653/v1/W19-1909.

\bibitem{adamw}
I.~Loshchilov and F.~Hutter, ``Decoupled weight decay regularization,'' in
  \emph{International Conference on Learning Representations}, 2019, doi:
  10.48550/arXiv.1711.05101.

\bibitem{prplot}
T.~Saito and M.~Rehmsmeier, ``The precision-recall plot is more informative
  than the {ROC} plot when evaluating binary classifiers on imbalanced
  datasets,'' \emph{PLOS ONE}, vol.~10, no.~3, p. e0118432, 2015, doi:
  10.1371/journal.pone.0118432.

\bibitem{calibration}
C.~Guo, G.~Pleiss, Y.~Sun, and K.~Q. Weinberger, ``On calibration of modern
  neural networks,'' in \emph{Proceedings of the 34th International Conference
  on Machine Learning}, vol.~70, 2017, pp. 1321--1330, doi:
  10.48550/arXiv.1706.04599.

\bibitem{bh}
Y.~Benjamini and Y.~Hochberg, ``Controlling the false discovery rate: A
  practical and powerful approach to multiple testing,'' \emph{Journal of the
  Royal Statistical Society Series B}, vol.~57, no.~1, pp. 289--300, 1995, doi:
  10.1111/j.2517-6161.1995.tb02031.x.

\end{thebibliography}

\end{document}